\documentclass[MSNbibl,nameyear,dvips]{arxstspdf}
\usepackage{flushend}
\usepackage{stfloats}

\volume{29}
\issue{3}
\pubyear{2014}
\firstpage{371}
\lastpage{374}
\doi{10.1214/14-STS491} 
\referstodoi{10.1214/14-STS480}
\docsubty{FLA}

\makeatletter

\renewcommand{\citep}[1]{(\citeauthor{#1}, \citeyear{#1})}
\makeatother

\begin{document}
\begin{frontmatter}

\vspace*{12pt}\title{Think Globally, Act Globally: An~Epidemiologist's Perspective on~Instrumental Variable Estimation}
\runtitle{An Epidemiologist's Perspective}

\begin{aug}
\author[a]{\fnms{Sonja A.} \snm{Swanson}\corref{}\ead[label=e1]{sswanson@hsph.harvard.edu}}
\and
\author[b]{\fnms{Miguel A.} \snm{Hern\'an}\ead[label=e2]{mhernan@hsph.harvard.edu}}
\runauthor{S. A. Swanson and M. A. Hern\'an}

\affiliation{Harvard School of Public Health}

\address[a]{Sonja A. Swanson is Post-Doctoral Research Fellow, Department of Epidemiology, Harvard School of Public
Health, Boston, Massachusetts 02115, USA \printead{e1}.}
\address[b]{Miguel A. Hern\'an is Professor, Department of Biostatistics and Epidemiology, Harvard School of Public
Health, Boston, Massachusetts 02115, USA
\printead{e2}.}
\end{aug}


\end{frontmatter}

We appreciated Imbens' summary and reflections on the state of
instrumental variable (IV) methods from an econometrician's
perspective. His review was much needed as it clarified several issues
that have been historically a source of confusion when individuals from
different disciplines discussed IV methods.

Among the many topics covered by Imbens, we would like to focus on the
common choice of the local average treatment effect (LATE) over the
``global'' average treatment effect (ATE) in IV analyses of
epidemiologic data. As Imbens acknowledges, this choice of the LATE as
an estimand has been contentious (\cite{angrist1996}; \cite{rg1996};
\cite{deaton2010}; \cite{imbens2009better}; \cite{pearl2011principal}). Several authors have
questioned the usefulness of the LATE for informing clinical practice
and policy decisions, because it only pertains to an unknown subset of
the population of interest: the so-called ``compliers.'' To make things
worse, many studies do not even report the expected proportion of
compliers in the study population \citep{swanson2013}. Other authors
have wondered whether the LATE is advocated for simply because of the
relatively weaker assumptions required for its identification,
analogous to the drunk who stays close to the lamp post and declares
whatever he finds under its light is what he was looking for all along
\citep{deaton2010}.

Here, we explore the limitations of the LATE in the context of
epidemiologic and public health research. First we discuss the
relevance of LATE as an effect measure and conclude that it is not our
primary choice. Second, we discuss the tenability of the monotonicity
condition and conclude that this assumption is not a plausible one in
many common settings. Finally, we propose further alternatives to the
LATE, beyond those discussed by Imbens, that refocus on the global ATE
in the population of interest.

\section{Relevance of a Local Average Treatment Effect in Epidemiologic
Research}

Some authors claim the LATE is actually what we are primarily
interested in, even if the ``compliers'' are not identifiable. A common
argument is that we care about the treatment effect for the
``compliers'' because this is the only subset of the population whose
treatment behaviors are modifiable. This rationale is problematic,
however, as the definition of ``compliers'' is instrument-dependent
\citep{pearl2011principal}. If multiple instruments were separately
used to estimate the effect of treatment in the ``compliers'' in the
same study, each effect estimate would be pertinent to a different
subset of the population: the ``compliers'' are different for each IV
analysis. It is unclear why the effects in all these various subsets
would be of primary interest. The perception of the ``compliers'' being
the subset whose behaviors are modifiable is overly simplistic because
it ignores this instrument dependence.

Other authors, like Imbens in his review, perceive the LATE as a
``second choice'' estimand, yet advocate it can sometimes be useful. He
argues for reporting subgroup effects even if the subgroup-specific
analysis is not exactly addressing the primary research question. He
proposes an analogy between estimating the effect in the ``compliers''
and estimating an effect in an all-male randomized trial, where males
are, like ``compliers,'' a subset of the general population. This
analogy begs the question: why would we be interested in the effect
estimate from a male-only trial? There are two possible reasons: (1) we
wish to inform clinical or policy decisions for men only, or (2) we
wish to extrapolate from the study to inform decisions for men and
women. If the former, the analogy with the ``compliers'' seems
ill-placed: as we\vadjust{\goodbreak} do not know who is a~``complier,'' we would not know
to whom our new policy should apply. If the latter, then we would need
to assume effect homogeneity between men and women. However, in IV
analyses, the LATE is often chosen over the global ATE precisely
because we expect too much effect heterogeneity for the ATE to be
validly identified. As such, extrapolation of the LATE to the entire
population could be ill-advised.

Finally, the LATE does not naturally translate to time-varying
treatments. Because many if not most exposures studied in epidemiologic
research vary over time, we cannot rely on the LATE to meaningfully
study their effects. If we want to study the effects of time-varying
treatments or exposures within the IV framework, we may instead
consider g-estimation of structural nested models. This approach
requires detailed modeling assumptions about the effect of treatment
\citep{robins2009estimation}.

\section{Plausibility of Monotonicity in Epidemiologic Research}

Part of the argument for favoring the LATE is that the requisite
monotonicity assumption appears more reasonable than the homogeneity
assumptions required to estimate the ``global'' ATE. For dichotomous
treatments and instruments, monotonicity requires no ``defiers'' exist,
while homogeneity requires there is no effect modification by the
instrument among the treated and untreated \citep{robins1989}. However,
while it may be plausible that there are essentially zero ``defiers''
in a randomized trial, the monotonicity condition may not hold for
instruments used in observational studies.

Consider one of the most commonly proposed instruments in epidemiologic
research, physician or facility prescribing preference \citep
{swanson2013}. Suppose we are interested in estimating the effect of a
specific treatment relative to no treatment among patients attending a
clinic where two physicians with different preferences work. The first
physician usually prefers to prescribe the treatment, but she makes
exceptions for her patients with diabetes (because of some known
contraindications). The second usually prefers to not prescribe the
treatment, but he makes exceptions for his more physically active
patients (because of some perceived benefits). Any patient who was both
physically active and diabetic would have been treated contrary to both
of these physicians preferences and, therefore, would be a ``defier.''
Because physicians' preferences represent the weighing of a variety of
risks and benefits, there may be many opportunities for a patient to be
treated contrary to physicians' preferences, and thus exhibit a
violation of monotonicity \citep{swansonmono}.

Moreover, the compliance types (``compliers,'' ``defiers,''
``always-takers,'' ``never-takers'') are not well-defined for such
instruments. Our example above considers a study with only two
physicians that could possibly have seen our patients. In more common
research settings with multiple physicians, for the compliance types to
be well-defined, all physicians with the same level of preference who
could have seen a patient would have to then treat the patient in the
exact same way. Because this is unrealistic, not only is it more likely
that there are monotonicity violations but whoever the ``compliers''
are that our effect pertains to is not just an unidentifiable but an
ill-defined subset of our population \citep{swansonmono}.

Further, most of the commonly proposed instruments in epidemiologic
research use a noncausal proxy instrument in their analyses. This is
done out of necessity, for example, we cannot measure the actual
preference of the physician when using a preference-based instrument,
or we sometimes only have the means to measure approximate locations in
the genome when using a genetic-based instrument. Although the use of
such a noncausal instrument could satisfy the other identifying
assumptions, this measurement error complicates our interpretation of a
LATE-like effect \citep{hr2006}. In particular, if the unmeasured
causal instrument is continuous, then the standard IV estimator using a
dichotomous proxy instrument would not be an effect in a specific
``compliant'' subpopulation but rather identifies a weighted average of
everybody with weights that are not particularly meaningful to any
policy decision. This is assuming that monotonicity held for the
unmeasured causal instrument, which is unlikely for instruments like
preference where the instrument is a summary of multiple dimensions of
encouragement \citep{swansonmono}.

\section{Alternative Approaches: A~Refocus on the Global Average
Treatment Effect}

Because the LATE is not generally relevant to epidemiologic research
questions, and the apparently plausible monotonicity assumption is
actually implausible in many common settings, we suggest shifting focus
back to the effect of primary interest, which is often the global ATE
\citep{rg1996}. Imbens summarized two options for this using IV
methods: (1)~present bounds for the ATE \citep{bp1997}, which are often
too wide to directly inform the particular decision at hand, or (2)
present a point estimate for the ATE assuming effect homogeneity \citep
{robins1994}, even though this assumption often is not palatable. Of
course this dichotomy is somewhat artificial: we can always do both.
Moreover, there are middle grounds.

Consider the canonical flu vaccine trial that\break  Imbens described:
physicians were randomized\break  to either receive or not receive a letter
encouraging~influenza vaccinations for their patients, and we are
interested in the effect of vaccination on flu-re\-lated hospitalizations
(\citeauthor{mcdonald1992},  \citeyear{mcdonald1992}). Under the instrumental conditions but not
monotonicity, Imbens calculated the Balke--Pearl bounds of [$-$0.24,
0.64] for the global ATE. These bounds do not allow us to conclude
whether vaccines are incredibly helpful, harmful, or somewhere in
between. If we further assume effect homogeneity, the point estimate is
$-$0.12 using the standard IV estimator that assumes additive
homogeneity. However, these homogeneity
assumptions are often perceived as too strong. Next, we propose a
middle ground between the uninformative bounds based on reasonable
assumptions (at least in the flu vaccine trial) and the point estimate
based on the often heroic assumption of homogeneity.

One reason the Balke--Pearl bounds are often wide is because (by
definition) we have no information on what would have happened to the
always-takers had they not been vaccinated and what would have happened
to the never-takers had they been vaccinated. The bounds are estimated
under the most extreme\linebreak[4]  scenarios where all or none of these patients
would be hospitalized under these unobserved counterfactual treatments.
However, we could use subject-matter knowledge to assume a more
reasonable range of possibilities. For example, we might propose that
at most 10\% of the never-takers under treatment and 10\% of the
always-takers under no treatment would be hospitalized. We can then use
extensions of the Balke--Pearl bounds to estimate bounds of [$-$0.07,
0.02] that are consistent with this further constraint and monotonicity
\citep{rr2010}. If our narrower bounds are correct, the estimated LATE
using the standard IV estimator under monotonicity ($-$0.12) overstates
the benefit of vaccination that would have occurred had we vaccinated
the whole population. If we assume stricter limits on what would have
happened to the never-takers under treatment (e.g., at most 5\% would
have been hospitalized), we can narrow the bounds and identify the
direction of the effect: [$-$0.07, $-$0.02]. A disadvantage of this
approach is that, like approaches based on estimating the LATE, it
requires well-defined compliance types, an assumption that may be
reasonable for this randomized trial but less appropriate in other
settings as we detailed above. For a review of other approaches to
partial identification of the global ATE under IV-type assumptions, see
\citet{swansonbds}.

Another middle ground approach is to describe the sensitivity of the
point estimate to the suspected effect heterogeneity. A problem with
this approach is that the homogeneity condition is mathematically
stated with respect to the instrument, which is not intuitive, and thus
makes it difficult to apply subject-matter knowledge toward
understanding the validity of the condition. To solve this problem,
\citet{hr2006} proposed a sufficient condition for heterogeneity that
is stated with respect to the confounders. This sufficient condition
allows us to use subject matter knowledge to understand its
plausibility---and, therefore, we can also propose sensitivity
analyses based on plausible violations of this assumption. An advantage
of this approach is we no longer need to assume the compliance types
are well-defined or of a known distribution. Some authors have
previously proposed ways to understand the implications of measured
effect modifiers \citep{brookhart2007preference}, and these ideas could
be extended to consider unmeasured effect modifiers as well.

\section{Conclusion}

Imbens states we are ``limited in the questions we can answer credibly
and precisely.'' We agree, but there are differences between the
questions we can answer and the questions we want answered. Choosing
only answerable questions (e.g., identifying the LATE in some settings)
can mislead decision-making efforts: our estimates may be
misinterpreted as directly relevant to a decision when in fact they are
only tangentially related. On the other hand, exact answers for our
questions (e.g., identifying the global ATE) may be often unattainable,
but a combination of data and assumptions based on subject-matter
knowledge may go a long way towards partly answering them (e.g.,
obtaining narrow bounds for the ATE). At the very least, incomplete
answers should serve as a reminder and encouragement that further
studies---using other data and/or other assumptions---are warranted.

\section*{Acknowledgments}
This work was supported in part by the National Institutes of Health
(R01 AI073127).



\end{document}